\documentclass{raa}% referee version: for submission
\pdfoutput=1
  \usepackage{graphicx,times}%for PS/EPS graphics inclusion, new
  \usepackage{natbib}
  \usepackage{amssymb,amsmath}
  \usepackage{lineno}
  
  \usepackage[pdftex=true,pagebackref=true]{hyperref}
  \usepackage{ulem}

  \newcommand{\Teff}{$T_\mathrm{eff}$}
  \newcommand{\logg}{log $\mathnormal{g}$}
  
  \newcommand{\loggf}{log $\mathnormal{gf}$}
  \newcommand{\vsini}{$v\,\mathrm{sin\,}i$}

  \newcommand{\Msolar}{M$_{\odot}$}

\begin{document}

%\linenumbers
%%%%%%%%%%%%%%%%%%%%%%%%%%%%%%%%%%%%%%%%%%%%%%%%%%%%%%%
%%%% Title, Author and Address
\title{LAMOST/HRS Spectroscopic Analysis of Two New Li-rich Giants}
%\subtitle{I. Place Your Subtitle Here}

\volnopage{Vol.0 (20xx) No.0, 000--000}%%preserved for Editor. DOn't remove!
\setcounter{page}{1}%%starting page, preserved for Editor. DOn't remove!

\author{Ze-Ming Zhou \inst{1,2}
\and Jianrong Shi \inst{1,2}
\and Hong-Liang Yan \inst{1,2}
\and Yong-Hui Hou \inst{2,3}
\and Kai Zhang \inst{3}
\and \\Qi Gao \inst{1,2}
\and Xiao-Dong Xu \inst{1,2}
\and Hai-Long Yuan \inst{1}
\and Yu-Tao Zhou \inst{4,5}
\and Kaike Pan \inst{6,7}
\and Zi-Ye Sang \inst{1}
\and Yong-Heng Zhao \inst{1}
}

\institute{CAS Key Laboratory of Optical Astronomy, National Astronomical Observatories, Chinese Academy of Sciences, Beijing 100101, China; {\it sjr@bao.ac.cn, hlyan@nao.cas.cn}\\
\and School of Astronomy and Space Science, University of Chinese Academy of Sciences, Beijing 100049, China\\
\and Nanjing Institute of Astronomical Optics \& Technology, National Astronomical Observatories, Chinese Academy of Sciences, Nanjing 210042, China\\
\and Department of Astronomy, Peking University, Beijing 100871, China\\
\and Kavli Institute for Astronomy and Astrophysics, Peking University, Beijing 100871, China\\
\and Apache Point Observatory, Sunspot, New Mexico 88349, USA\\
\and Astronomy Department, New Mexico State University, Las Cruces, New Mexico 88003, USA\\
\vs\no {\small Received~~20xx month day; accepted~~20xx~~month day}}

%%%%%%%%%%%%%%%%%%%%%%%%%%%%%%%%%%%%%%%%%%%%%%%%%%%%%%%
%%%% Abstract
\abstract{Two Li-rich candidates, TYC\,1338-1410-1 and TYC\,2825-596-1, were observed with the new high-resolution echelle spectrograph, LAMOST/HRS. Based on the high-resolution and high-signal-to-noise ratio (SNR) spectra, we derived stellar parameters and abundances of 14 important elements for the two candidates. The stellar parameters and lithium abundances indicate that they are Li-rich K-type giants, and they have A(Li)$_\mathrm{NLTE}$ of 1.77 and 2.91\,dex, respectively. Our analysis suggests that TYC\,1338-1410-1 is likely a red giant branch (RGB) star at the bump stage, while TYC\,2825-596-1 is most likely a core helium-burning red clump (RC) star. The line profiles of both spectra indicate that the two Li-rich giants are slow rotators and do not show infrared (IR) excess. We conclude that engulfment is not the lithium enrichment mechanism for either star. The enriched lithium of TYC\,1338-1410-1 could be created via Cameron-Fowler mechanism, while the lithium excess in TYC\,2825-596-1 could be associated with either non-canonical mixing processes or He-flash.
}

\authorrunning{Z.-M. Zhou, J.-R. Shi, H.-L. Yan, et al.}%author_head in even pages
\titlerunning{LAMOST/HRS Spectroscopic Analysis of Two New Li-rich Giants}% title_head in odd pages

\maketitle
%% The author head (on even pages) and the title head (on odd pages) will be
%% automatically extracted from \author{} and \title{}. Whenever the title is too long,
%% you will be asked to supply a shorter one by inserting either \authorrunning{} or
%% \titlerunning{} before \maketitle. Anyway, you can specify your own heads.
%%
%%
%% Note: In the following text body of your manuscript, please note several differences from
%%       other major journals:
%% (1) \subsection{Please Capitalize the First Letter of Each Notional Word in Subsection Title}
%% (2) Please Capitalize the First Letter of Each Notional Word in all tables' captions

%%%%%%%%%%%%%%%%%%%%%%%%%%%%%%%%%%%%%%%%%%%%%%%%%%%%%%%
%%%% Introduction
\section{Introduction} \label{sec:introduction}
It is expected that the surface abundance of lithium in a K giant would be low because its surface lithium is diluted during stellar evolution from the main sequence to the red giant branch \citep[RGB,][]{1989ApJS...71..293B, 2000A&A...359..563C}. During the first dredge-up (FDU) process, surface materials, including lithium, are transported to the interior regions. And lithium is depleted due to high temperature there. Standard models predict that the lithium abundance A(Li)\footnote{A(Li) = log[n(Li)/n(H)] + 12} in a low-mass solar metallicity star is no more than $\sim$ 1.5\,dex after the FDU process \citep{2010A&A...522A..10C, 2020A&A...633A..34C}. However, observations have revealed that some giants have lithium abundances higher than 1.5\,dex during the last three decades \citep[e.g.,][]{1989ApJS...71..293B, 2011A&A...529A..90M, 2011ApJ...730L..12K, 2012ApJ...752L..16K, 2013MNRAS.430..611M, 2014A&A...569A..55A, 2016MNRAS.461.3336C}, and a few of them have lithium abundances even higher than that ($\sim$ 3.3\,dex\footnote{3.3\,dex is the A(Li) in meteorites \citep[CI carbonaceous chondrites,][]{2009ARA&A..47..481A} and therefore is inferred as the ISM lithium abundance at the solar metallicity.}) of the interstellar medium (ISM) \citep[e.g.,][]{1995ApJ...448L..41D, 2000ApJ...542..978B, 2005AJ....129.2831R, 2013MNRAS.430..611M, 2018A&A...615A..74Z, 2018NatAs...2..790Y}. \cite{1989ApJS...71..293B} found that about 1\% of giants show surface lithium enrichment based on a sample of 644 field giants. Most recently, \cite{2019ApJS..245...33G} showed that nearly 1.29\% of giants exhibit lithium excess based on a sample of more than 814,000 giants selected from the Large Sky Area Multi-Object Fiber Spectroscopy Telescope (LAMOST) survey.

Various scenarios have been conceived aiming at expounding ``anomalous" Li-rich phenomena in giants. The external mechanisms include pollution by planets/brown dwarfs \citep[e.g.,][]{1967Obs....87..238A, 1989ApJS...71..293B, 1989A&A...215...66G}, nova ejecta \citep[e.g.,][]{1994ApJ...435..791M} and mass transfer from stellar companions \citep[e.g.,][]{2016ApJ...819..135K}. Proposed internal processes are mostly related with the fresh lithium \citep[e.g.,][]{1999ApJ...510..217S, 2000A&A...359..563C} produced by the Cameron-Fowler mechanism \citep{1971ApJ...164..111C}.

The detailed information of Li-rich giants can help us better understand how and where lithium nucleosynthesis be triggered, and can constrain the lithium evolution history in our Galaxy. Therefore, the Li-rich objects have attracted attentions of researchers since the first source is confirmed. Obviously, it is very helpful to identify more Li-rich samples for better understanding the enrichment mechanisms. Since its pilot survey \citep{2012RAA....12.1197C, 2012RAA....12..723Z}, LAMOST has obtained more than 10 million low-resolution spectra. There is no doubt that such a massive dataset will serve as a great resource for searching Li-rich candidates \citep[e.g.,][]{2018ApJ...852L..31L, 2018NatAs...2..790Y, 2018A&A...615A..74Z, 2019ApJ...877..104Z, 2019MNRAS.482.3822S, 2019ApJS..245...33G}. And high-resolution spectral data are needed to confirm whether candidates are real Li-rich stars. Recently, a high-resolution spectrograph, LAMOST/HRS, was in commission. This instrument is a single-object and fiber-fed spectrograph with a resolution power of R\,$\sim$\,30,000, and its wavelength coverage is from 3,800 to 7,300\,{\AA}. It is fed by one of the thirty-five fibers distributed at different positions in LAMOST's field-of-view of 5 degrees in diameter. LAMOST/HRS provides a flexible functionality to perform high-resolution observations parallel to low/medium spectra surveys. This will enable us to obtain high-resolution and high-signal-to-noise ratio (SNR) spectra of Li-rich giants to better understand the lithium evolutionary history, together with the parallactic information from $Gaia$ \citep{2018A&A...616A...1G}.

In this study, we analyze the spectra of two Li-rich candidates taken with LAMOST/HRS. The paper is organized as follows. In Section \ref{sec:data}, we describe the observations and data reductions of our sample stars. Stellar parameters, abundances, and evolutionary statuses are determined in Section \ref{sec:analysis}. In Section \ref{sec:discussion}, we discuss possible mechanisms of lithium enhancement for the two stars. A brief summary is presented in the last section.

%%%%%%%%%%%%%%%%%%%%%%%%%%%%%%%%%%%%%%%%%%%%%%%%%%%%%%%
%%%% Observations
\section{Observations and Data Reduction} \label{sec:data}
A newly developed high-resolution spectrograph, LAMOST/HRS, has recently been commissioned. We select two Li-rich candidates, TYC\,1338-1410-1 and TYC\,2825-596-1, from low-resolution ($R \sim$ 2,000) spectra of the LAMOST DR5\footnote{http://dr5.lamost.org}. The two candidates are selected by template matching of the lithium resonance line at 6708\,{\AA} as described in \cite{2019ApJS..245...33G}.

\begin{figure}
   \centering
   \includegraphics[scale=0.34]{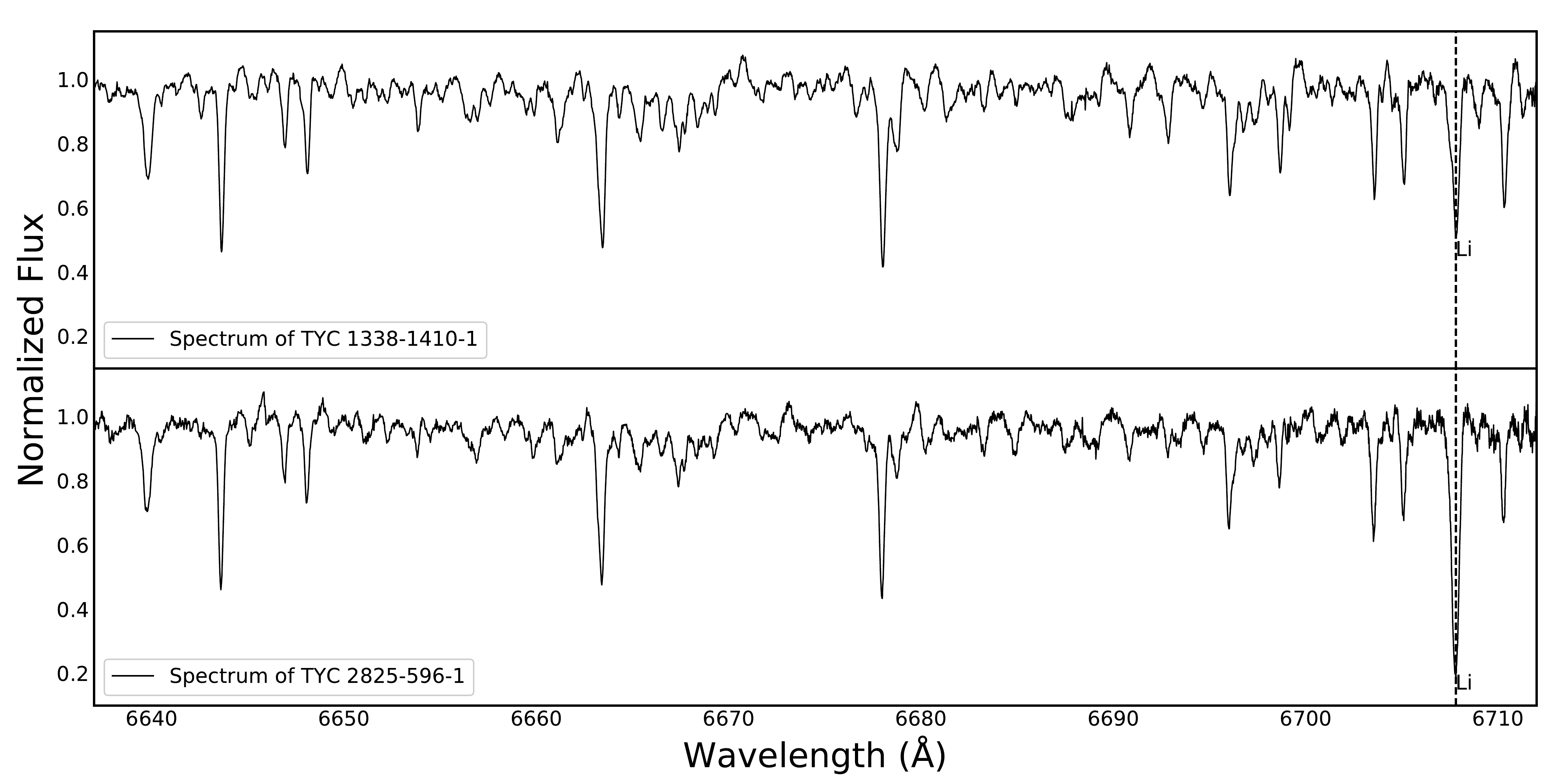}
   \caption{The sample spectra are shown for the two stars. The dash line indicates the lithium resonance line $\lambda$6708\,{\AA}.}
   \label{fig:SpectraSample}
\end{figure}

The followup observations are carried out in Nov. 2018 with LAMOST/HRS at the Xinglong Observatory (China). Six 30-minute exposures are obtained for TYC\,1338-1410-1, and five for TYC\,2825-596-1 during two bright nights. The coordinates of the two stars, along with their other properties, are given in Table \ref{tab:information}. The spectra are reduced with the IDL codes \citep{1998A&AS..130..381P} in a typical fashion: cosmic ray removal, bias level subtraction, flat field correction, spectral extraction, wavelength calibration, and co-adding individual exposures. The final SNRs of the spectra around $\lambda$5500\,{\AA} for both stars are higher than 100. Fig. \ref{fig:SpectraSample} shows the spectral section of the targets in the region of lithium line at 6708\,{\AA}.

\begin{table}
  \bc
  \begin{minipage}[]{100mm}
   \caption[]{Stellar Key Information}\label{tab:information}
  \end{minipage}
  \doublerulesep 0.5pt \tabcolsep 25pt %space between two columns
  \small
  \begin{tabular}{lrr}
    \hline\noalign{\smallskip}
      & TYC\,1338-1410-1& TYC\,2825-596-1\\\hline
    RA & 6:45:28.27  & 1:29:58.27 \\
    Dec. & +19:07:35.00  & +44:45:36.57 \\
    $V$(mag) & 9.34 $\pm$ 0.02 & 9.82 $\pm$ 0.03 \\
    $K$(mag) & 6.25 $\pm$ 0.04 & 7.05 $\pm$ 0.02 \\
    ${Gaia}$ DR2 ID & 3359663576603514240 & 397016434561325312 \\
    %\textbf{Parallax$_{Gaia}$ (mas)}&  2.82$\pm$0.07  &  1.19$\pm$0.06 \\
    Parallax (mas)& 2.82 $\pm$ 0.07 & 1.19 $\pm$ 0.06 \\
    Mass ($\mathrm{M_{\odot}}$) & 1.18 $\pm$ 0.05 & 1.58 $\pm$ 0.05 \\
    {\Teff}$_{,\ V-K}$ (K) & 4471 $\pm$ 100 & 4552 $\pm$ 100 \\
    {\logg}$_{pho}$ (dex) & 2.52 $\pm$ 0.20 & 2.18 $\pm$ 0.20 \\
    log($L/L_{\odot}$) & 1.62 $\pm$ 0.30 & 2.08 $\pm$ 0.30 \\
    {\Teff} (K) & 4652 $\pm$ 80 & 4631 $\pm$ 80 \\
    {\logg} (dex) & 2.37 $\pm$ 0.20 & 2.15 $\pm$ 0.20 \\
    {[Fe/H]} (dex) & -0.10 $\pm$ 0.10 & -0.20 $\pm$ 0.10 \\
    $\xi_\mathrm{t}$ (km\,s$^{-1}$) & 1.40 $\pm$ 0.20 & 1.40 $\pm$ 0.20 \\
    {\vsini} (km\,s$^{-1}$) & 2.27 $\pm$ 0.31 & 1.85 $\pm$ 0.40 \\
    \noalign{\smallskip}\hline
  \end{tabular}
   \ec
   \tablecomments{0.8\textwidth}{The $VK$ magnitudes are from VizieR. The stellar masses are estimated from stellar evolutionary tracks with corresponding stellar metallicity.}
\end{table}

%%%%%%%%%%%%%%%%%%%%%%%%%%%%%%%%%%%%%%%%%%%%%%%%%%%%%%%%%%%
%%%% Data analyses
\section{Data analyses} \label{sec:analysis}

%%%% Stellar Parameters
\subsection{Stellar Parameters} \label{sec:parameters}
The stellar parameters (effective temperature {\Teff}, surface gravity {\logg}, metallicity [Fe/H] and microturbulent velocity $\xi_\mathrm{t}$) of the two stars are derived through a spectroscopic method. A set of well-calibrated Fe I and Fe II lines from \cite{2018NatAs...2..790Y}, which was collected from \cite{2002PASJ...54.1041T}, \cite{2011A&A...528A..87M} and \cite{2012ApJ...757..109C}, is  adopted. The criteria for selecting lines have been described by \cite{2018NatAs...2..790Y}. Briefly, the selected Fe I and Fe II lines should be neither too strong ($>$\,120\,m{\AA}) nor too weak ($<$\,20\,m{\AA}). Because the spectral qualities near the Fe I lines at 4630\,{\AA} and 6750\,{\AA} of TYC\,1338-1410-1 are poor, and the core of the Fe II line at 5264\,{\AA} of TYC\,2825-596-1 is affected by a cosmic ray hit, these lines are excluded from our spectroscopic analysis. 34 Fe I and 7 Fe II lines are used in the analysis for TYC\,1338-1410-1, while 36 Fe I and 6 Fe II lines for TYC\,2825-596-1. Before determining the iron abundances for our two sample stars, we firstly adjust {\loggf} values to ensure the solar iron abundances derived from each line equal to 7.5\,dex. The effective temperatures are derived by requesting the iron abundances from Fe I lines being independent of line excitation potentials, and the surface gravities are determined by forcing ionization equilibrium between Fe I and Fe II lines. The micro-turbulent velocities are obtained by requesting the iron abundances from individual Fe I lines being independent of their reduced equivalent widths (EWs). We use the 1D MARCS-OS model atmospheres \citep{2008A&A...486..951G} in the process of determining stellar parameters. The derived stellar parameters shown in Table \ref{tab:information} indicate that both sample stars are K-type giants.

To cross check stellar parameters determined with the spectroscopic method, we also derive the photometric effective temperatures and surface gravities for both sample stars. The photometric data of the targets are collected from the fourth U.S. Naval Observatory CCD Astrograph Catalog \cite[UCAC4,][]{2013AJ....145...44Z} on the VizieR\footnote{http://vizier.u-strasbg.fr/viz-bin/VizieR} website. The photometric effective temperatures are obtained by using the calibration from \cite{1999A&AS..140..261A}. Their method is based on empirical relations, and slightly large uncertainties are found for objects of temperatures lower than 4000\,K and higher than 5500\,K compared with the theoretical calibrations. Thus, this calibration is relatively suitable for the two stars in this work. The photometric effective temperatures are 4471\,K and 4552\,K for TYC\,1338-1410-1 and TYC\,2825-596-1, respectively. The difference between spectroscopic and photometric effective temperatures is 79\,K for TYC\,2825-596-1 and 181\,K for TYC\,1338-1410-1. With the parallaxes taken from $Gaia$ DR2 \citep{2018A&A...616A...1G}, we derive the photometric surface gravities for our sample stars using the formula:

\begin{equation}
   \mathrm{log\ \mathnormal{g} = log\ \mathnormal{g_{\odot}} +log(\frac{M\ \ }{M_{\odot}}) +4log(\frac{\mathnormal{T}_{eff\ \ }}{\mathnormal{T}_{eff\odot}}) + 0.4(\mathnormal{M}_{bol} - \mathnormal{M}_{bol\odot})} \tag {1}
   \label{eq:logg}
\end{equation}

\noindent Here, M and $\mathrm{M_{\odot}}$ are the stellar and solar masses, respectively. The stellar mass is estimated by the position of a star in the Hertzsprung-Russell (HR) diagram with corresponding metallicity (Z) and masses of theoretical stellar evolution tracks. The theoretical stellar evolution tracks are taken from the PAdova and TRieste Stellar Evolution Code \citep[PARSEC\footnote{\tt https://people.sissa.it/\%7Esbressan/parsec.html},][]{2012MNRAS.427..127B, 2013EPJWC..4303001B}. $\mathrm{\mathnormal{M}_{bol}}$ and $\mathrm{\mathnormal{M}_{bol\odot}}$ denote stellar and solar absolute $V$ bolometric magnitudes, respectively. The adopted solar values are {\logg$_{\odot}$} = 4.44, {\Teff$_{\odot}$} = 5,777\,K, $M\mathrm{_{bol\odot}}$ = 4.74\,mag. The stellar absolute bolometric magnitudes are computed with the relation:

\begin{align}
    M_{\mathrm{bol}} & = \mathnormal{BC_V} + \mathnormal{M_V} \nonumber\\
                     & = \mathnormal{BC_V} + (m_V - A_V - (5\mathrm{log}\,d - 5)) \nonumber\\
                     & = \mathnormal{BC_V} + m_V + 5\mathrm{log}\,{\pi} + 5.0 - {A}_{V} \tag{2}
    \label{eq:Mbol}
\end{align}

\noindent Here, $BC_V$ is the bolometric correction in $V$ magnitude, and $\pi$ is the stellar parallax with a unit of arc-second. The bolometric corrections are calculated using the empirical relation from \cite{1999A&AS..140..261A}, and calculated $BC_V$ values for TYC\,1338-1410-1 and TYC\,2825-596-1 are -0.40 and -0.41, respectively. $A_\mathnormal{V}$ is the interstellar extinction in the direction of stellar sources, which we take from the Galactic Dust Reddening and Extinction\footnote{https://irsa.ipac.caltech.edu/applications/DUST/} \citep[and references therein]{2011ApJ...737..103S}. The adopted $A_\mathnormal{V}$ values for TYC\,1338-1410-1 and TYC\,2825-596-1 are 0.4838 and 0.2385 mag, respectively. We obtain a $M_{\mathrm{bol}}$ value of 0.70 for TYC\,1338-1410-1 and -0.45 for TYC\,2825-596-1. The fundamental parameters of the sample stars derived with both spectroscopic and photometric methods are tabulated in Table \ref{tab:information}. As one can see that the parameters from the two methods are in reasonable agreement.

The rotational velocity ({\vsini}) is a useful indicator to distinguish different lithium enrichment mechanisms. The rotational velocities of our giants have been derived by fitting the Fe I lines at 6726, 6733, and 6750\,{\AA} with the procedures described in \cite{2012ApJ...757..109C}. We choose these Fe I lines as they are relatively clean and locate near the lithium line at 6708\,{\AA}. The instrumental broadening is estimated from nearby Th-Ar wavelength calibration lines, and a Gaussian broadening is adopted for the instrumental profile fitting. In this process, we use the averaged FWHM of Th-Ar lines near the selected individual Fe I lines as the width of instrumental profile. The macroturbulent velocities are estimated by adopting the relations from \cite{2007A&A...475.1003H}, and they are 2.78 and 2.73 \,km\,s$^{-1}$ for TYC\,1338-1410-1 and TYC\,2825-596-1, respectively. With the estimated instrumental broadenings and macroturbulent velocities, we obtain rotational velocities of 2.27\,km\,s$^{-1}$ for TYC\,1338-1410-1 and 1.85\,km\,s$^{-1}$ for TYC\,2825-596-1, which suggests that neither star is a rapid rotator compared with normal K giants.

%%%% Abundances Determination
\subsection{Abundances Determination} \label{sec:abundances}
\begin{figure}
   \centering
   \includegraphics[scale=0.24]{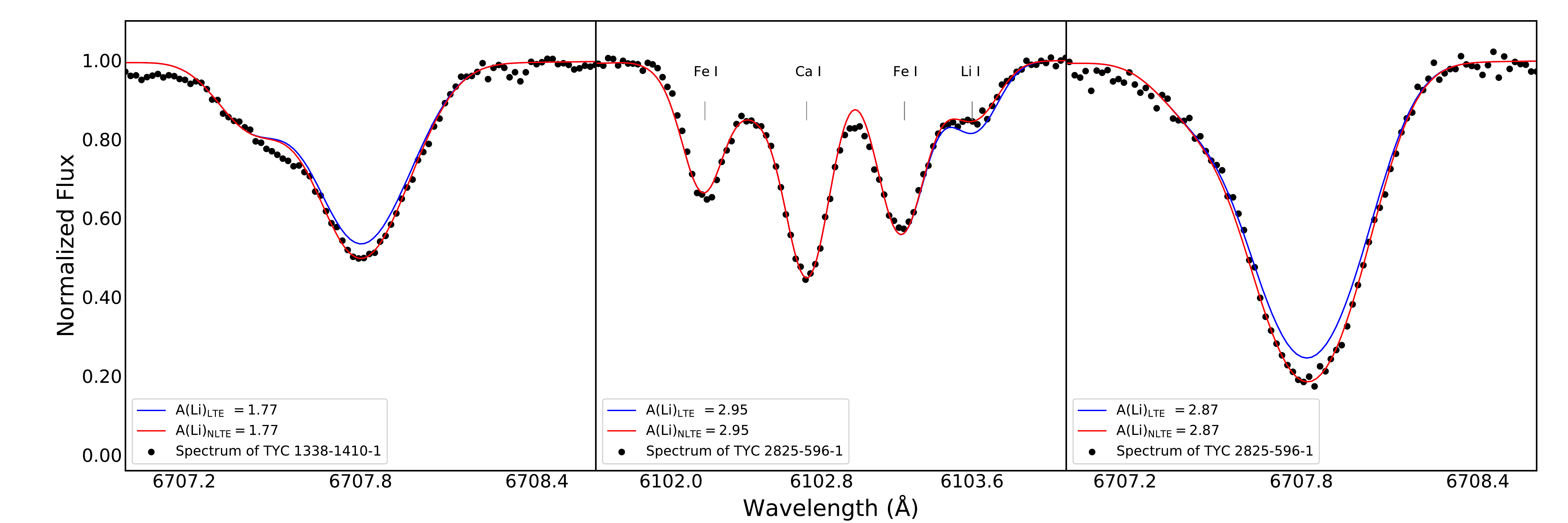}
   \caption{The left panel shows the synthetic spectra (solid lines) around lithium line at 6708\,{\AA} compared with observed data (black dots) for TYC\,1338-1410-1. The middle and right panels show corresponding spectral syntheses around lithium lines at 6104 and 6708\,{\AA} for TYC\,2825-596-1. The blue and red solid lines represent LTE and NLTE synthetic spectra,  respectively. The A(Li) values of LTE are plotted the same as those of corresponding NLTE syntheses to illustrate the NLTE effects.}
   \label{fig:LiFitting}
\end{figure}

The elemental abundance pattern is important to help us understand the enrichment mechanism of lithium. We derive abundances  of several important elements for our sample stars. The IDL/Fortran-based software, SIU, of \cite{1991ReetzDiploma} is used to determine the elemental abundances with a spectral synthesis method. The synthetic lines are calculated using the MARCS-OS model atmospheres \citep{2008A&A...486..951G} with the stellar parameters derived from the spectroscopic manner. The resonance doublet line at 6708\,{\AA} and the subordinate line at 6104\,{\AA} have been included to derive the lithium abundances, and the atomic line data are taken from \cite{2007A&A...465..587S}. Both the two lines are used to derive lithium abundances for TYC\,2825-596-1, but only the resonance line is used for TYC\,1338-1410-1 because its subordinate line is relatively weak. In Fig. \ref{fig:LiFitting}, we present the fitting results. For the resonance lithium line at 6708\,{\AA}, the theoretical LTE line profiles do not show any sign of saturation for our program stars, so both the LTE and NLTE synthetic spectra can well reproduce the observed data. It is worthy to note that the LTE line profiles in Fig \ref{fig:LiFitting} are computed with the lithium abundances determined from the NLTE fittings for the corresponding stars in  order to show the NLTE effect on each line.

\begin{table}
   \bc
   \begin{minipage}[]{100mm}
    \caption[]{Chemical Abundances [X/Fe]}\label{tab:abundances}
   \end{minipage}
   \doublerulesep 0.5pt \tabcolsep 25pt %space between two columns
   \small
   \begin{tabular}{lrr}
     \hline\noalign{\smallskip}
     Species                                 & TYC\,1338-1410-1 & TYC\,2825-596-1   \\\hline
     A(Li)$\mathrm{^{6104}_{LTE}}$ &              ... & 2.85 $\pm$ 0.07 \\
     A(Li)$\mathrm{^{6104}_{NLTE}}$&              ... & 2.95 $\pm$ 0.07 \\
     A(Li)$\mathrm{^{6708}_{LTE}}$ & 1.85 $\pm$ 0.13& 3.10 $\pm$ 0.15 \\
     A(Li)$\mathrm{^{6708}_{NLTE}}$& 1.77 $\pm$ 0.12& 2.87 $\pm$ 0.14 \\
     $\mathrm{[C\ I/Fe]_{LTE}}$  &-0.12 (1)         &-0.12 (1)          \\
     $\mathrm{[O\ I/Fe]_{LTE}}$  & 0.21 (1)         & 0.15 (1)          \\
     $\mathrm{[Na\ I/Fe]_{LTE}}$ & 0.24 $\pm$ 0.04 (4)& 0.19 $\pm$ 0.04 (4) \\
     $\mathrm{[Na\ I/Fe]_{NLTE}}$& 0.11 $\pm$ 0.03 (4)& 0.06 $\pm$ 0.04 (4) \\
     $\mathrm{[Mg\ I/Fe]_{LTE}}$ & 0.24 $\pm$ 0.04 (5)& 0.15 $\pm$ 0.05 (4) \\
     $\mathrm{[Al\ I/Fe]_{LTE}}$ & 0.18 (1)           & 0.13 $\pm$ 0.03 (2) \\
     $\mathrm{[Si\ I/Fe]_{LTE}}$ & 0.16 $\pm$ 0.06 (6)& 0.12 $\pm$ 0.04 (8) \\
     $\mathrm{[Ca\ I/Fe]_{LTE}}$ & 0.02 $\pm$ 0.07 (13)& 0.02 $\pm$ 0.05 (13) \\
     $\mathrm{[Ca\ I/Fe]_{NLTE}}$& 0.05 $\pm$ 0.05 (13)& 0.08 $\pm$ 0.05 (13) \\
     $\mathrm{[Sc\ II/Fe]_{LTE}}$& 0.10 $\pm$ 0.05 (5)& 0.04 $\pm$ 0.10 (5) \\
     $\mathrm{[Ti\ II/Fe]_{LTE}}$& 0.05 $\pm$ 0.04 (9)&-0.01 $\pm$ 0.07 (8) \\
     $\mathrm{[Cu\ I/Fe]_{LTE}}$ &-0.02 $\pm$ 0.07 (3)&-0.06 $\pm$ 0.04 (2) \\
     $\mathrm{[Cu\ I/Fe]_{NLTE}}$&-0.02 $\pm$ 0.06 (3)&-0.06 $\pm$ 0.05 (2) \\
     %$\mathrm{[Sr\ II/Fe]_{LTE}}$ &-0.09&... \\
     $\mathrm{[Zr\ II/Fe]_{LTE}}$& 0.00 (1)         & 0.03 (1)\\
     $\mathrm{[Ba\ II/Fe]_{LTE}}$& 0.04 $\pm$ 0.02 (3)&-0.02 $\pm$ 0.04 (3) \\
     $\mathrm{[Eu\ II/Fe]_{LTE}}$& 0.12 (1)         & 0.09 (1) \\
      \noalign{\smallskip}\hline
   \end{tabular}
    \ec
    \tablecomments{0.8\textwidth}{The number in brackets is the count of lines used for respective elemental abundance. Uncertainties for lithium lines are due to the alteration in {\Teff}, and uncertainties for other species are abundance standard deviations.}
\end{table}

Abundances of several other important elements are also derived where possible. Most of our line data are adopted from \cite{2016ApJ...833..225Z}, while the {\loggf} values from \cite{2014A&A...568A..25N} and the Vienna Atomic Line Database (VALD) are used for O I forbidden line at 6300.304 {\AA} and Eu II line at 6645.064 {\AA}. For most of these species, we derive only LTE abundances except for sodium, calcium, and copper. All the derived abundances are presented in Table \ref{tab:abundances}. The lithium abundances indicate that both our sample stars are indeed Li-rich.

It is well known that lithium abundance is very sensitive to {\Teff} \citep[e.g.,][]{2007A&A...465..587S}, so testing how much A(Li) will change with the uncertainty of {\Teff} is important. Our fittings show that A(Li)$\mathrm{^{6708}_{NLTE}}$ increases 0.12\,dex for TYC\,1338-1410-1, A(Li)$\mathrm{^{6104}_{NLTE}}$ and A(Li)$\mathrm{^{6708}_{NLTE}}$ increase 0.07 and 0.14\,dex for TYC\,2825-596-1, respectively, if {\Teff} raises 80\,K.

\begin{figure}
   \centering
   \includegraphics[scale=0.48]{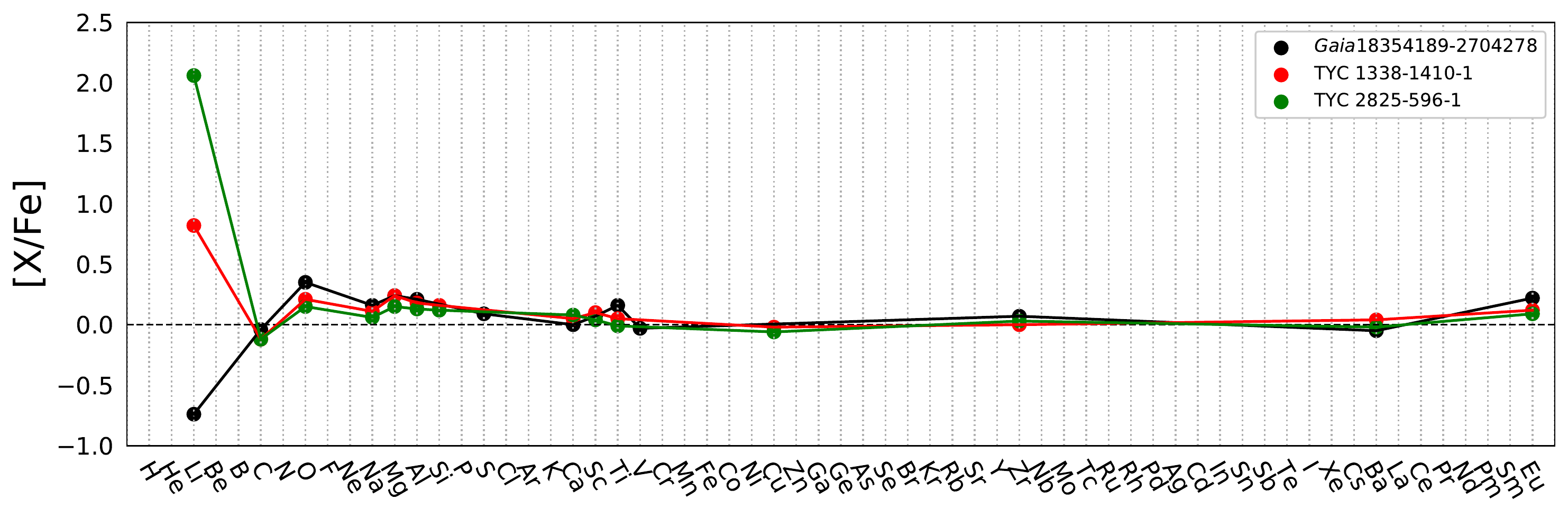}
   \caption{Element-to-element comparisons between the Li-normal star $Gaia$18354189-2704278 and our two Li-rich giants. The abundances of lithium, sodium, calcium, and copper are plotted using the NLTE values. The corresponding ionized abundances from the $Gaia$-ESO DR3 catalog are used for elements which have highly-ionized species.}
   \label{fig:AtomAbun}
\end{figure}

An element-to-element abundance comparison can potentially be helpful in identifying nucleosynthesis processes in a star, and finding possible origin of the lithium enrichment. Therefore, we compare the elemental abundances of our Li-rich giants with those of Li-normal giants. For this purpose, we select the Li-normal star $Gaia$18354189-2704278, as a comparison target. This star has similar stellar parameters with our sample stars, and its parameters are {\Teff} = 4675\,K, {\logg} = 2.37\,dex, [Fe/H] = -0.19\,dex and $\xi_\mathrm{t}$ = 1.41\,km\,s$^{-1}$. Chemical abundances of the comparison star are taken from the DR3 catalog of $Gaia$-ESO\footnote{Choose and click the title, $Gaia$-ESO spectroscopic survey, on https://www.eso.org/qi/ to log in to the query page.} \citep{2013Msngr.154...47R}. In order to display its abundances in [X/Fe]\footnote{[X/Fe] = log$\mathrm{\frac{\Bigl(n(X)/n(Fe)\Bigr)_{*}}{\Bigl(n(X)/n(Fe)\Bigr)_{\odot}}}$ = log$\biggl(\frac{n(X)/n(H)}{n(Fe)/n(H)}\biggr)_{*}$ - log$\biggl(\frac{n(X)/n(H)}{n(Fe)/n(H)}\biggr)_{\odot}$ = [X/H] - [Fe/H]} form, we use the solar elemental abundances from \cite{2007SSRv..130..105G}. The element-to-element comparisons are shown in Fig. \ref{fig:AtomAbun}. The figure shows that the [Li/Fe] of our two Li-rich giants are about 2-3\,dex higher than that of the Li-normal star, while the abundances of other elements are comparable. This is consistent with the finding by \cite{2016MNRAS.461.3336C}.

%%%% Evolutionary Status
\subsection{Evolutionary Status} \label{sec:evolution}
Stellar evolutionary status is one of the key factors in understanding specific lithium enrichment mechanism of Li-rich giants. Thus, we determine the positions of our stars in their HR diagrams. First, we calculate their stellar luminosities by using:

\begin{equation}
    M\mathrm{_{bol} - \mathnormal{M}_{bol\odot} = -2.5log(\frac{\mathnormal{L\ }}{\mathnormal{L}_{\odot}})} \tag{3}
\end{equation}

\begin{figure}
   \centering
   \includegraphics[scale=0.50]{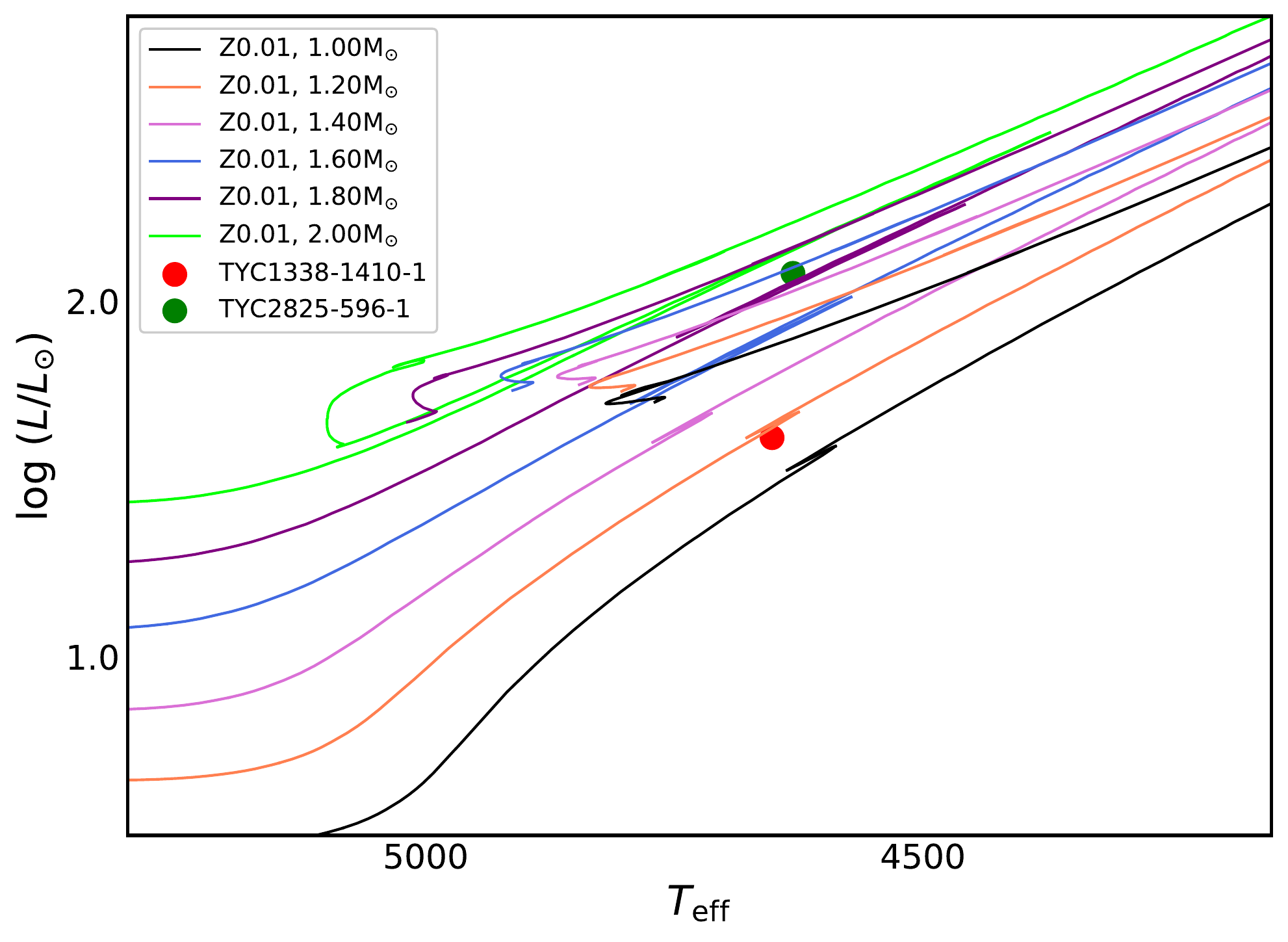}
   \caption{Locations of the Li-rich K giants TYC\,1338-1410-1 (red dot) and TYC\,2825-596-1 (blue dot) in the HR diagram of luminosity versus effective temperature according to PARSEC evolution tracks. The mass interval is from 1.0\,M$_\odot$ to 2.0\,M$_\odot$ in step of 0.2\,M$_\odot$ with the corresponding metallicity Z = 0.01.}
   \label{fig:eTrack}
\end{figure}

\noindent By adopting the $M\mathrm{_{bol}}$ values from Section \ref{sec:parameters}, we obtain log($L/L_{\odot}$) values of 1.62 and 2.08 for TYC\,1338-1410-1 and TYC\,2825-596-1, respectively. Fig. \ref{fig:eTrack} shows their positions in the HR diagram with stellar evolution tracks from PARSEC \citep{2012MNRAS.427..127B, 2013EPJWC..4303001B}. Their positions suggest that TYC\,1338-1410-1 is more likely an RGB bump giant, while it is hard to unambiguously distinguish between RC and RGB statuses for TYC\,2825-596-1 only by its location. \cite{2018ApJ...858L...7T} suggested that TYC\,2825-596-1 is an RC object based on a standard machine-learning approach.

%%%% IR Excess
\subsection{IR Excess} \label{sec:IRexcess}
\cite{1996ApJ...456L.115D} and \cite{1997ApJ...482L..77D} proposed a scenario that all normal K giants with masses from 1.0\,{\Msolar} to 2.5\,{\Msolar} become Li-rich for a short time ($\mathrm{\sim 10^{5}}$\,yr ) during their RGB stage, and lots of giants in this phase are associated with far-infrared excess, which is due to the expanding circumstellar gas and dust shells triggered by the internal mixing mechanism. \cite{1999MNRAS.304..925S, 1999MNRAS.308.1133S} proposed that engulfing of substellar companions could increase observed lithium abundances, and should also result in IR-excess. Thus, it is desired to investigate whether the two targets have IR excess. We adopt the Equation (1) from Section 5.4 of \cite{2015AJ....150..123R}:

\begin{equation}
    \mathrm{\chi_{[3.4],[22]} = \frac{([3.4]-[22])_{observed} - ([3.4]-[22])_{predicted}}{\sigma_{([3.4]-[22])}}} \tag{4}
\end{equation}

\begin{figure}
    \centering
    \includegraphics[scale=0.50]{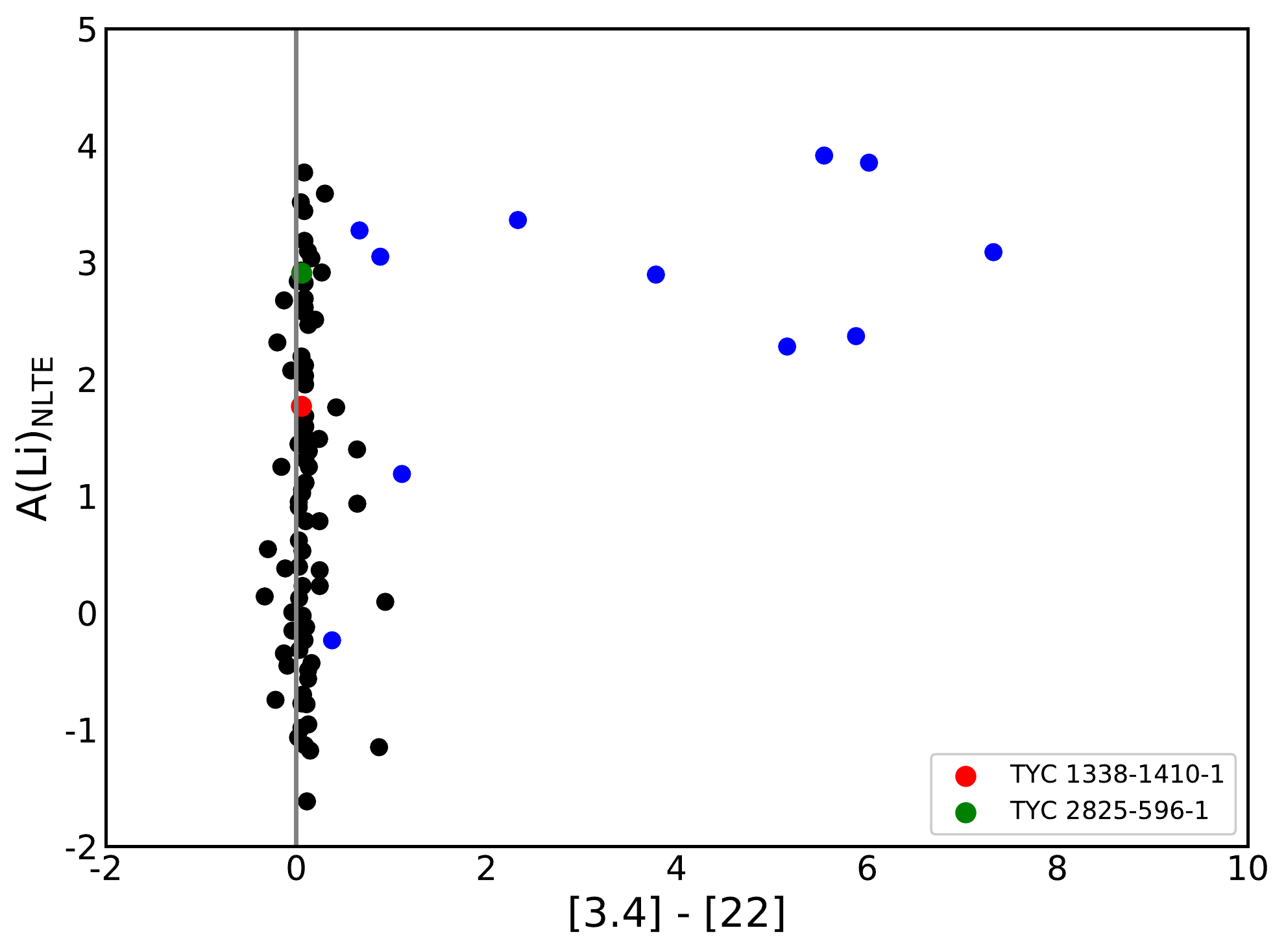}
    \caption{A(Li)$\mathrm{_{NLTE}}$ vs. [3.4] - [22]. The blue dots are stars with IR excess from \cite{2015AJ....150..123R}, and the black dots are stars without IR excess from \cite{2015AJ....150..123R}. The red and green dots are our giants.}
    \label{fig:IRexcess}
\end{figure}

\noindent Here, [3.4] and [22] denote the magnitudes of 3.4\,$\mu$m and 22\,$\mu$m, and $\sigma_{([3.4]-[22])}$ is the quadratic sum of the errors of 3.4\,$\mu$m and 22\,$\mu$m. Normally, stars with $\chi_{[3.4],[22]}$ $>$ 3.0 are considered as obvious IR-excess sources. ([3.4]-[22])$\mathrm{_{predicted}}$ is 0 for K giants \citep{2012AJ....144..135M}. Using observed data from the WISE online data catalog\footnote{http://vizier.cfa.harvard.edu/viz-bin/VizieR?-source=II/328} \citep{2003yCat.2246....0C}, we get $\mathrm{\chi_{[3.4],[22]}}$ values of 0.46 and 0.61 for TYC\,1338-1410-1 and TYC\,2825-596-1, respectively. This result indicates that both our stars do not show obvious IR excess. \cite{2015AJ....150..123R} systematically investigated possible correlations between lithium abundance and IR-excess for 176 cleanest giants, and found that the objects with large IR excess tend to have high lithium abundances, but most of Li-rich stars do not show IR excess in their Figure 18. The calculated IR excesses of our two stars do not contrast with their findings. Fig. \ref{fig:IRexcess} plots ([3.4]\,-\,[22]) versus A(Li)$\mathrm{_{NLTE}}$ for our samples, along with the 176 red giants from \cite{2015AJ....150..123R}, and it clearly shows that both of our sample stars are Li-rich stars without IR excess.

%%%%%%%%%%%%%%%%%%%%%%%%%%%%%%%%%%%%%%%%%%%%%%%%%%%%%%%%%%%
%%% Discussion
\section{Discussion} \label{sec:discussion}
Lithium enrichment mechanisms for RC giants remain unclear, although a large fraction, if not the majority of Li-rich stars, are found in their RC evolutionary stages \citep[e.g.,][]{2011ApJ...730L..12K, 2019ApJ...877..104Z, 2019ApJ...880..125C}. Several lithium enrichment scenarios have been proposed, for example, \cite{2011ApJ...730L..12K} suspected that the lithium production via the Cameron-Fowler mechanism may also occur at the core-helium flash stage. \cite{2019MNRAS.482.3822S} argued that the mechanism of the lithium enrichment phenomena in RC giants might associate with either helium flash at the RGB tip or substellar companion engulfments. \cite{2019ApJ...878L..21S} systematically investigated 24 Li-rich RC giants with asteroseismology-based data from $Kepler$, and suggested that the helium flash could be responsible for the observed lithium excess for their RC giants. Combining with spectroscopic and asteroseismic analyses, \cite{2014ApJ...784L..16S} confirmed that KIC\,5000307 is an RC Li-rich giant, and suggested that the most-likely mechanism for its lithium excess is due to the non-canonical mixing events at the RGB tip or during the helium flash. \cite{2019ApJ...880..125C} studied a stellar sample of 2330 Li-rich giant stars, selected from the LAMOST low-resolution DR2 spectra library, and found that near 80\% of the Li-rich giants are experiencing core-helium burning phase. They proposed that the tidal interactions and planetary engulfment are seemingly the most reasonable lithium enrichment mechanisms for their data. Very recently, \cite{2020ApJ...889...33Z} computed the merger evolution models of a helium-core white dwarf (HeWD) with an RGB star, and suggested that Li-rich RC stars can be formed through this process.

For RGB stars, fresh lithium can be produced via Cameron-Fowler mechanism \citep{1971ApJ...164..111C}. In this mechanism, $\mathrm{^{7}Be}$ is created in the deeper interior of a star, and then be rapidly transported to the region where the temperature is much cooler before the decay of $\mathrm{^{7}Be(e^{-}, \nu)^{7}Li}$ takes place. \cite{2000A&A...359..563C} suggested that the destruction of mean molecular weight barrier can lead to a short period of lithium production by extra mixing undergoing inside a low-mass RGB star. The extra mixing is a process that links the bottom of convective zone with stellar interior, where $\mathrm{^{7}Be}$ is newly formed to produce fresh lithium at stellar surface via Cameron-Fowler mechanism.

One external source to replenish lithium, as has been suggested by \cite{1967Obs....87..238A}, \cite{1999MNRAS.308.1133S}, \cite{2010ApJ...723L.103C, 2012ApJ...757..109C} and \cite{2016ApJ...829..127A, 2016ApJ...833L..24A}, is the engulfment of substellar companions, such as giant planets or brown dwarfs. \cite{1999MNRAS.304..925S, 1999MNRAS.308.1133S} investigated the possible observational signatures of lithium enrichment by accretion of planets/brown dwarfs for AGB and RGB stars. Their computations indicated that, along with the increasement of surface lithium, the host star will spin up because of the deposition of orbital angular momentum of the substellar companion, and eject a shell from the surface with a subsequent phase of IR excess. \cite{2010ApJ...723L.103C, 2012ApJ...757..109C} suggested that accretion of a fast close-orbiting giant planet can also accelerate the rotational velocity of the host RGB star due to the transfer of angular momentum. \cite{2016ApJ...833L..24A} modeled the engulfment of a substellar companion for giants with different masses and metallicities, and they suggested that there is an upper limit of A(Li)\,$\sim$\,2.2\,dex for the engulfment mechanism. One of our stars, TYC\,2825-596-1, has a lithium abundance significantly higher than this upper limit, so the engulfment mechanism, at least it alone, cannot explain the lithium enrichment for this star. There are many studies \citep[e.g.,][]{2002AJ....123.2703D} investigated correlation between lithium abundances and rotational velocities, for example, \cite{2000A&A...363..239D}, \cite{2002AJ....123.2703D} and \cite{2013AN....334..120C} found that giants with high rotational velocities are more likely to have enriched surface lithium abundances. This is not a surprise  because the engulfment increases observed lithium abundance and rotational velocity of a star whenever the mechanism applies. Considering both of our stars have low rotational velocities and high lithium abundances, it is unlikely that the engulfment is the main mechanism of lithium enrichment for our stars.

It is suggested that the lithium enrichment by engulfment will not only spin up the host star but also result in IR excess. Both our sample stars have normal rotational velocities with {\vsini}\,$<$\,8\,km\,s$^{-1}$, and do not show IR excess. Our results suggest that the enhanced lithium of TYC\,1338-1410-1 is from the Cameron-Fowler mechanism, and the excess lithium of TYC\,2825-596-1 is associated with either the non-canonical mixing or He-flash. It should be pointed out that additional information is needed to better constrain the physical characteristics of our two stars, and to fully understand their lithium enhancement mechanisms.

%%%%%%%%%%%%%%%%%%%%%%%%%%%%%%%%%%%%%%%%%%%%%%%%%%%%%%%%%%%
%%%% Summary
\section{Summary}
A newly developed high-resolution spectrograph, LAMOST/HRS, has recently been commissioned. We have observed two Li-rich candidates, TYC\,1338-1410-1 and TYC\,2825-596-1, with this new instrument. The two candidates are selected from the LAMOST DR5 low-resolution spectral library, and they are comfirmed as Li-rich K giants. The main results can be summarized as follows.
\begin{itemize}
   \item The NLTE lithium abundances are 1.77 and 2.91\,dex for TYC\,1338-1410-1 and TYC\,2825-596-1, respectively. The abundances of other 13 important elements for both objects have also been derived, and their abundance patterns are similar to the Li-normal giant except lithium.
   \item Their masses and evolutionary statuses have been obtained. It is found that TYC\,1338-1410-1 is more likely an RGB star with a mass of 1.18\,{\Msolar}, while TYC\,2825-596-1 is most likely an RC giant with a mass of 1.58\,{\Msolar}.
   \item The lithium enrichment for both stars do not favour the engulfment mechanism, as they are slow rotators, and do not have IR excess. TYC\,1338-1410-1 is more likely to enhance its surface lithium via the Cameron-Fowler mechanism, while the enrichment mechanism of the RC star TYC\,2825-596-1 is less clear. Possible mechanisms include the non-canonical mixing during core-helium burning phase, helium flash, and the hypothesis of the merge of an RGB star with a HeWD.
\end{itemize}

%%%%%%%%%%%%%%%%%%%%%%%%%%%%%%%%%%%%%%%%%%%%%%%%%%%%%%%%%%%
%%% ORCIDs
\section{ORCID iDs}
Ze-Ming Zhou: \url{https:/orcid.org/0000-0002-1619-1660}\\
Jianrong Shi: \url{https:/orcid.org/0000-0002-0349-7839}\\
Hong-Liang Yan: \url{https:/orcid.org/0000-0002-8609-3599}\\
%Yong-Hui Hou: \url{}\\
Qi Gao: \url{https://orcid.org/0000-0003-4972-0677}\\
Xiao-Dong Xu: \url{https:/orcid.org/0000-0003-4789-0621}\\
Hai-Long Yuan: \url{https:/orcid.org/0000-0002-4554-5579}\\
Yu-Tao Zhou: \url{https://orcid.org/0000-0002-4391-2822}\\
Kaike Pan: \url{https:/orcid.org/0000-0002-2835-2556}\\
%Zi-Ye Sang: \url{}

%%%%%%%%%%%%%%%%%%%%%%%%%%%%%%%%%%%%%%%%%%%%%%%%%%%%%%%%%%%
%%%% Acknowledgements
\begin{acknowledgements}
We thank the referees for their pertinent comments and suggestions that helped us to improve the quality of our manuscript. Our research is supported by National Key R\&D Program of China No.2019YFA0405502, the Key Research Program of the Chinese Academy of Sciences under grant No.XDPB09-02, the National Natural Science Foundation of China under grant Nos. 11973052, 11833006, 11473033, 11603037, and the International partnership program's Key foreign cooperation project, Bureau of International Cooperation, Chinese Academy of Sciences under grant No.114A32KYSB20160049. This work is supported by the Astronomical Big Data Joint Research Center, co-founded by the National Astronomical Observatories, Chinese Academy of Sciences and Alibaba Cloud. This work is also partially supported by the Open Project Program of the Key Laboratory of Optical Astronomy, National Astronomical Observatories, Chinese Academy of Sciences. Z.-M.Z. thanks for the financial support from China Scholarship Council (CSC, No.201604910642) for his study at New Mexico State University in the United States of America and would like to express the gratitude to all those who have helped during the writing of this thesis. H.-L.Y. acknowledges the supports from Youth Innovation Promotion Association, Chinese Academy of Sciences. Guoshoujing Telescope (the Large Sky Area Multi-Object Fiber Spectroscopic Telescope LAMOST) is a National Major Scientific Project built by the Chinese Academy of Sciences. Funding for the project has been provided by the National Development and Reform Commission. LAMOST is operated and managed by the National Astronomical Observatories, Chinese Academy of Sciences. This research has made use of the NASA/IPAC Infrared Science Archive, which is funded by the National Aeronautics and Space Administration and operated by the California Institute of Technology. This publication makes use of data products from the Wide-field Infrared Survey Explorer, which is a joint project of the University of California, Los Angeles, and the Jet Propulsion Laboratory/California Institute of Technology, funded by the National Aeronautics and Space Administration. This work is based on observations collected at the European Southern Observatory under ESO programme ID 188.B-3002 and has also made use of abundant data collected from the NIST and VALD databases.
\end{acknowledgements}

%%%%%%%%%%%%%%%%%%%%%%%%%%%%%%%%%%%%%%%%%%%%%%%%%%%%%%%
%%%% bibliography
\bibliographystyle{raa.bst}
\bibliography{ms2020-0158.bib}

\label{lastpage}

\end{document}